\documentclass{rspublic}
\newcommand{\unit}[1]{\ensuremath{\, \mathrm{#1}}}
\usepackage{graphicx}%,amsmath,amssymb,setspace,anysize,cite}
\begin{document}
\title[Dynamics of concentrated microgel systems]{Multiple dynamic regimes in concentrated microgel systems}
\author[D. A. Sessoms and others]{David A. Sessoms$^{1}$, Irmgard Bischofberger$^{1}$, Luca Cipelletti$^{2}$, and V\'{e}ronique Trappe$^{1}$}
\affiliation{1. Department of Physics, University of Fribourg, Switzerland \\ 2. LCVN UMR5587, University of Montpellier 2 and CNRS, France}
 \label{firstpage} \maketitle

\begin{abstract}{microgels, soft colloids, glasses, jamming, heterogeneous dynamics, photon correlation imaging}

We investigate dynamical heterogeneities in the collective
relaxation of a concentrated microgel system, for which the packing
fraction can be conveniently varied by changing the temperature. The
packing fraction dependent mechanical properties are characterised
by a fluid-solid transition, where the system properties switch from
a viscous to an elastic low-frequency behaviour. Approaching this
transition from below, we find that the range $\xi$ of spatial
correlations in the dynamics increases. Beyond this transition,
$\xi$ reaches a maximum, extending over the entire observable system
size of $\sim 5 \unit {mm}$. Increasing the packing fraction even
further leads to a second transition, which is characterised by the
development of large zones of lower and higher dynamical activity
that are well separated from each other; the range of correlation
decreases at this point. This striking non-monotonic dependence of
$\xi$ on volume fraction is reminiscent of the behaviour recently
observed at the jamming/rigidity transition in granular systems
(Lechenault \textit{et al.} 2008). We identify this second
transition as the transition to `squeezed' states, where the
constituents of the system start to exert direct contact forces on
each other, such that the dynamics becomes increasingly determined
by imbalanced stresses. Evidence of this transition is also found in
the frequency dependence of the storage and loss moduli, which
become increasingly coupled as direct friction between the particles
starts to contribute to the dissipative losses within the system. To
our knowledge, our data provide the first observation of a
qualitative change in dynamical heterogeneity as the dynamics switch
from purely thermally-driven to stress-driven.

\end{abstract}

\section{Introduction}

Soft deformable colloids, such as foam bubbles, emulsion droplets,
star polymers and microgels can be efficiently packed beyond the
close packing conditions of hard spheres. At such high packing
fraction, the constituents of the system touch each other, exerting
direct forces on each other, which results in their deformation.
Consequently we expect the dynamics in such a `squeezed' state to be
mainly determined by force balances. Any imbalance in the forces,
i.e. stresses, will lead to motion and thus to reconfiguration of
the system. Indeed, we can presume that the dynamics of deformable
spheres is predominantly determined by either stress-imbalances or
temperature depending on their packing fractions, and the goal of
this work is to assess some of the differences between stress-driven
and thermally-driven dynamics. Our investigation is motivated by the
assumption that the high packing fraction behaviour of repulsive
deformable spheres is determined by two critical conditions: a) the glass
transition at which the particles become so densely packed that the
diffusion of the particles past their nearest neighbours becomes
highly improbable on an experimental time scale and b) the jamming
transition at which the particles are jammed together, forming a
network of direct contacts. That both transitions may be significant
can be inferred from the combined datasets obtained by respectively
Meeker \textit{et al.} (1997) and Mason \textit{et al.} (1995).

\begin{figure}[tb]
\centering
\includegraphics[width=0.6\textwidth]{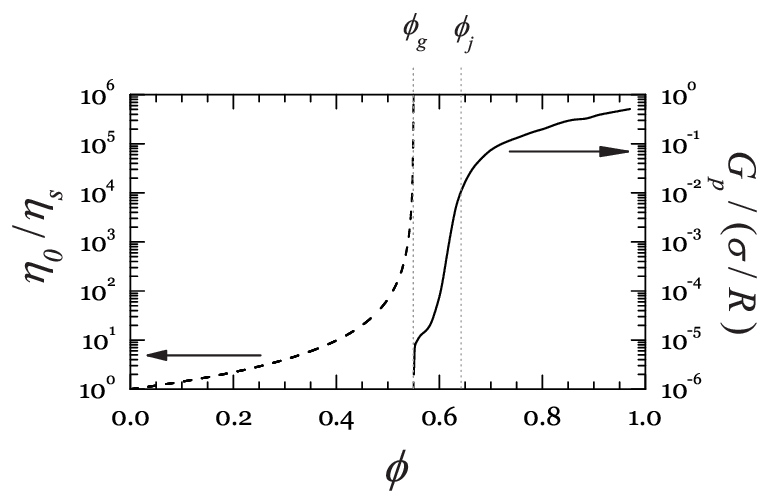}
\caption[Volume fraction dependence of the mechanical properties of
hard and soft spheres] { Volume fraction dependence of the
mechanical properties of deformable spheres inferred from data sets
obtained for hard spheres in the lower volume fraction range (dashed
line) (Meeker \textit{et al.} 1997) and for emulsions in the higher
volume fraction range (solid line) (Mason \textit{et al.}~1995). The
low shear viscosity $\eta_{0}$ seemingly diverges at the glass
transition volume fraction, $\phi_{g}$, which is also characterised
by the onset of low-frequency elasticity $G_{p}$. Beyond $\phi_{g}$,
$G_{p}$ increases quickly towards the jamming transition (random
close packing) $\phi_{j}$. Beyond $\phi_{j}$, the elasticity
increases nearly linearly with $(\phi-\phi_{j})$.  The low shear
viscosity is normalized by the solvent viscosity $\eta_{s}$, while
the modulus is normalized by the modulus intrinsic to the emulsion
droplet, $\sigma/R$, with $\sigma$ the surface tension and $R$ the
radius of the drop.} \label{fig:PuseyMason}
\end{figure}

Meeker \textit{et al.} (1997) investigated the low shear viscosity
$\eta_{0}$ of hard sphere colloidal systems. According to their
investigations, the low shear viscosity exhibits an apparent divergence at the
experimental glass transition $\phi_{g} = 0.58$, where they find
that an unconstrained fit of the Krieger-Dougherty equation,
$\eta_{0}/\eta_{s} = \left(1-\phi/\phi_{c}\right)^{-\beta}$, to
their data yield the parameters $\phi_{c} = 0.55$ and $\beta=1.8$,
where $\eta_{s}$ is the viscosity of the background medium; these
fit results are shown as a dashed line in fig.~\ref{fig:PuseyMason}.
Mason \textit{et al.} (1995) investigated the elasticity of dense
emulsions beyond $\phi_{c} = 0.55$; the volume fraction dependence
of the elastic modulus $G_{p}$ normalized by the ratio of the
surface tension $\sigma$ to droplet radius $R$ is shown as a solid
line in fig.~\ref{fig:PuseyMason}. According to their
investigations, the volume fraction dependence is determined by both
the experimental glass transition at $\phi_{g}$ and the jamming
transition at $\phi_{j} \sim 0.64$.  For $\phi_{g}<\phi<\phi_{j}$,
the elasticity increases quickly with increasing volume fraction,
while for $\phi>\phi_{j}$ the elasticity increases approximately linearly
with $\phi-\phi_{j}$. Reflecting the squeezed nature of the system
at very high $\phi$, the elasticity converges towards the elasticity
intrinsic to emulsion droplets, i.e. $\sigma/R$, for $\phi$
approaching unity (Princen 1983). Inferring that the emulsion
droplets would behave like the hard sphere system below the glass
transition, we therefore expect three states for a deformable
sphere: at $\phi < \phi_{g}$ the system behaves like a supercooled
fluid, at $\phi_{g}<\phi<\phi_{j}$ the system is in a glassy state,
at $\phi>\phi_{j}$ the system is in a squeezed state.

In this paper we present our work on the dynamics of thermosensitive
microgels, as an example of deformable spheres, and we show that these
three states are characterised by a distinct
behaviour of dynamical heterogeneity. More precisely, spatial
correlations of the dynamics are relatively short-ranged and of
small amplitude in the supercooled state, grow dramatically in the
glassy state, where they span the whole system, and finally become
more localized, but of large amplitude, in the squeezed state.

\section{Sample characteristics: conformational and mechanical properties}

We use aqueous solutions of poly-N-isopropylacrylamide (PNiPAM)
microgels, which exhibit a lower critical solution temperature
(LCST) at $\sim33\unit{^\circ C}$ (Senff \& Richtering 1999).  Below
the LCST, the dimensions of our PNiPAM microgels exhibit a
remarkable sensitivity to temperature, which enables us to vary the
volume fraction of a given system by a factor of 1.7 by varying the
temperature from $20\unit{^\circ C}$ to $30\unit{^\circ C}$.  Though
very convenient for the study of the high volume fraction behaviour,
working with microgel solutions calls for some caution, if the aim
is to compare their behaviour to other soft sphere systems like foams
and emulsions.  Indeed, microgel solutions are solutions and not
dispersions; they are in thermodynamic equilibrium with the solvent
and do not possess a well defined interface. They thus can partly
interpenetrate and/or compress under their own osmotic pressure.
Despite this enhanced difficulty, their unique temperature
sensitivity bears two main advantages: the volume fraction can be
varied by changing the temperature and, more importantly, the sample
history can be controlled uniquely by temperature, with no need to
apply a preshear to the sample, as is typically required by
colloidal systems. Indeed, the study of the dynamics of highly
concentrated systems, in particular of glassy and squeezed systems,
can tremendously suffer from an ill-controlled history of the
sample. The thermosensitive microgel systems offer the unique
advantage to allow for the preparation of the system in a fluid-like
state at low volume fraction (high temperature), which is
subsequently quenched to the solid-like states by inflating the
particle (decreasing the temperature), so as to obtain a higher
volume fraction. This protocol is not only convenient to use in almost any
experimental set-up, it is to a certain extent more meaningful than
the application of a high shear stress or rate, which is generally
used to erase the sample history (Cloitre \textit{et al.} 2000;
Rogers \textit{et al.} 2008). Indeed, the application of shear
imposes a directional deformation to the system, while the isotropic
inflation of the particles results in randomly oriented stresses
i.e. strains.

Our PNiPAM-microgels are synthesized as described by Senff \&
Richtering (1999) and purified by extensive dialysis against water.
We subsequently produce a highly concentrated stock solution by
evaporating water at a temperature of $60^\circ \unit{C}$ and
reduced pressure using a rotary evaporator.  This stock solution is
then used to produce more dilute samples. The actual concentration
of the stock is determined by drying a defined solution volume and
determining the residual amount of microgels to obtain both a
concentration in weight\% and a concentration in g/ml, where the
latter is defined relative to the solution volume at a room
temperature of $\sim 20^\circ \unit{C}$. Due to the use of an ionic
initiator, our microgels are charged.  To screen  these charges, we
add a solution of sodium-thiocyanate, NaSCN, to all our PNiPAM
solutions, so that the final salt concentration in our systems is
$0.03\unit{M}$.

We characterise the temperature dependent conformational properties of
our microgels at a concentration of $4\cdot 10^{-5}\unit{g/ml}$ by
static and dynamic light scattering using a commercial light
scattering apparatus. We characterise both salt-free and salted
microgel solutions to ascertain that the salt does not significantly
change the solubility of the microgels.  Both systems exhibit essentially the same temperature dependence for  $T$ smaller than the lower critical solution temperature $T_{c}$. Increasing the temperature
leads to a gradual decrease of the particle size up to $T_{c}$, where we find that the temperature dependence of the hydrodynamic radius $R_{h}$ and the radius of gyration $R_{g}$ is well approximated by a critical-like function of form $R_{g,h}=A\left(T_{c}-T\right)^{a}$. The resulting fit parameters for $R_{h}$ are $T_{c}=33.4\unit{^\circ C}$, $A=86.3\unit{nm}$, and $a=0.131$; for $R_{g}$, we find $T_{c}=33.7\unit{^\circ C}$, $A=46.9\unit{nm}$, and $a = 0.191$. In agreement with previous investigations on similar PNiPAM-microgels, the hydrodynamic radius is significantly larger than the radius of gyration for $T < T_{c}$ (Senff \& Richtering 2000; Clara Rahola 2007). This is due to
the uneven distribution of cross-linkers, the cross-linking density
being larger in the centre of the microgels than on the edges (Mason
\& Lin 2005). Because of the higher density of the core, the core scatters significantly more than the lower density edges, such that we essentially measure the radius of the highly cross-linked core in static light scattering, while we capture the hydrodynamically effective radius of the particle in dynamic light scattering. As the temperature approaches the critical solution temperature, the overall density of the microgel increases, such that $R_{h}$ approaches $R_{g}$.  Beyond the LCST, the characterisation of the microgel dimensions is only possible for the salt-free solution,
where the unscreened charges of the particles prevent aggregation.
In this range of temperature we find that $R_{g}/R_{h} \cong 0.75$,
near the value expected for hard sphere colloids (Dhont
1996).

Near the LCST, a change in temperature affects both intra- and
inter-molecular interactions of the microgels (Wu \textit{et al.}
2003). To assess the range of temperatures at which a change in temperature will only inflate or deflate the
particle but will not significantly alter the inter-particle
interactions, we determine the temperature dependence of the low
shear viscosity $\eta_{0}$ of microgel solutions in the range of
concentrations of $1.15\cdotp 10^{-3} \unit{g/ml}$ to $4.63\cdotp
10^{-2}\unit{g/ml}$. Our rheological measurements are performed in
the lower concentration range ($1.15\cdotp 10^{-3} \unit{g/ml}$ to
$1.15\cdotp 10^{-2} \unit{g/ml}$) with an Ubbelohde viscometer;
in the higher concentration range ($8.64 \cdotp
10^{-3}\unit{g/ml}$ to $4.63\cdotp 10^{-2}\unit{g/ml}$) we use a
commercial stress-strain rheometer equipped
with a cone and plate geometry (cone and plate radius
$R=25\unit{mm}$, cone angle $\alpha=1.0^\circ$).  To ensure the best
possible temperature control and to avoid evaporation in the
rheometer, we use a temperature hood combined with a solvent trap
(Sato \& Breedveld 2005). This guarantees that our samples will not exhibit evaporation-induced changes over an experimental time window of
2--3 hours. At higher concentrations, the history of our samples is
carefully controlled using the following protocol: the samples are
equilibrated in a fluid state at $32\unit{^\circ C}$, where we additionally apply a shear rate of $\dot{\gamma} = 1000 \unit{s^{-1}}$ to fully erase any previous quench history. The temperature is then decreased to the desired temperature and equilibrated at this
temperature for $500\unit{s}$ prior to any experiment, where we have
tested that our systems do not exhibit any significant changes in
their rheological properties beyond this waiting time.

The low shear viscosity is determined in shear rate experiments,
where we ramp the shear rate from $\dot{\gamma} =
10^{-5}\unit{s^{-1}}$ -- $10^{3}\unit{s^{-1}}$, depending on
temperature and microgel concentration.  For each shear rate, we
maintain $\dot{\gamma}$ for a sufficiently long time to allow for
the system to reach steady state.  The steady state viscosity
$\eta\left(\dot{\gamma}\right)$ exhibits a shear rate dependence
that is typical of colloidal suspensions.  At low concentrations and
high temperature, we find that $\eta\left(\dot{\gamma}\right)$
exhibits almost no dependence on $\dot{\gamma}$; in these cases we
determine $\eta_{0}$ as the mean of $\eta\left(\dot{\gamma}\right)$,
$\eta_{0} = \left\langle
\eta\left(\dot{\gamma}\right)\right\rangle$. At higher
concentrations and lower temperatures the system exhibit shear
thinning behaviour; in these cases we obtain $\eta_{0}$ by fitting
$\eta\left(\dot{\gamma}\right)$ to the Cross-equation,
\begin{equation}
  \eta\left(\dot{\gamma}\right) = \eta_{\infty} + \frac{\eta_{0}-\eta_{\infty}}{1+\left(C\dot{\gamma}\right)^{m}}
    \label{eq:Cross}
\end{equation}
where $\eta_{\infty}$ is the high shear viscosity and $1/C$ is a
measure of the crossover shear rate denoting the onset of shear
thinning.

\begin{figure}[tb]
\centering
\includegraphics[width=1.0\textwidth]{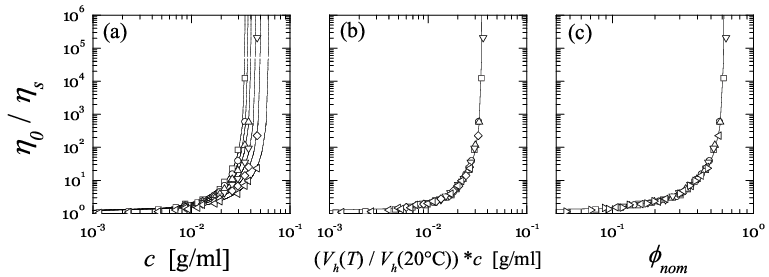}
\caption[Concentration dependence of the low shear viscosity of
microgel solutions] {(a) Concentration dependence of the relative
low shear viscosity $\eta_{0}/\eta_{s}$ of microgel solutions at
temperatures ranging from 20$\unit{^\circ C}$--30$\unit{^\circ C}$;
from left to right: $T=$ 20, 22, 24, 26, 28, $30^\circ \unit{C}$.
(b) Renormalizing the concentration by the degree of shrinkage of
the microgel particles with temperature results in a single master
curve. (c)  Assuming a voluminosity of $k = 17.6 \unit{ml/g}$ at
$T=20\unit{^\circ C}$,  $c$ is converted into a nominal volume
fraction $\phi_{nom}$. All lines denote critical-like divergences.}
\label{fig:ViscScale}
\end{figure}

In fig.~\ref{fig:ViscScale}(a) we report the relative viscosity
$\eta_{0}/\eta_{s}$ as a function of concentration $c$. The measurements are performed at temperatures ranging from $20\unit{^\circ C}$ to $30\unit{^\circ
C}$. Our data exhibit distinct variations as the temperature is
changed. At any given temperature, we find that $\eta_{0}/\eta_{s}$
increases dramatically with concentration, exhibiting a
critical-like divergence at some critical concentration.  This
critical concentration systematically shifts to higher values as the
temperature is increased.  That this behaviour is entirely due to
the change in the particle volume is shown in
fig.~\ref{fig:ViscScale}(b).  A simple renormalisation of the
concentration by the ratio $V_{h}\left(T\right) / V_{h}\left(20^\circ
\unit{C}\right)$ collapses the data onto a single master
curve, where we use $V_{h}\left(20^\circ
\unit{C}\right) \approx R^{3}_{h}$ as a reference volume, such that the
concentration axis refers to the one obtained at $20\unit{^\circ
C}$.  This almost perfect collapse demonstrates the equivalence
between varying the volume fraction by changing the particle
concentration and varying the volume fraction by changing the
particle volume via temperature (Senff \& Richtering 1999, 2000). For our microgels, this equivalence
holds up to $T=30\unit{^\circ C}$; for temperatures larger than
$T=30\unit{^\circ C}$, the concentration dependence of the viscosity
exhibits deviations from the scaling behaviour, reflecting a change
in the particle-particle interactions. Our master-curve is well
approximated by a critical-like function,
$\eta_{0}/\eta_{s}=\left(1-c/c_{c}\right)^{\alpha}$, with a
critical concentration of $c_{c} \cdotp
V_{h}\left(T\right)/V_{h}\left(20^\circ \unit{C}\right)=0.035\unit{g/ml}$ and a critical exponent of $\alpha=2.25$.  
To express concentrations in terms of volume fractions, we
assume a voluminosity of $k = 17.6 \unit{ml/g}$ for our systems at
$T=20\unit{^\circ C}$, which we use to define a nominal volume
fraction according to $\phi_{nom}= k\cdotp c\cdotp
V_{h}\left(T\right)/V_{h}\left(20^\circ \unit{C}\right)$. This voluminosity is
chosen based on different mapping techniques, including the one
where the low concentration dependence of $\eta_{0}/\eta_{s}$ is
mapped to the behaviour expected for hard spheres (Senff \&
Richtering 1999, 2000). However, we find that different approaches
can lead to a variation in $k$ of almost 20\%, such that the
indicated nominal volume fraction $\phi_{nom}$ can only be regarded
as an approximate gauge of the volume fraction. Nonetheless, for
convenience we use $\phi_{nom}$ instead of $c \cdotp
V_{h}\left(T\right)/V_{h}\left(20^\circ \unit{C}\right)$ to indicate our
concentrations. For the concentration and temperature dependence of
$\eta_{0}/\eta_{s}$ this results in the dependence shown in
fig.~\ref{fig:ViscScale}(c), where the critical divergence according
to $\eta_{0}/\eta_{s}=\left(1-\phi_{nom}/\phi_{nom,c}\right)^{\alpha}$
occurs at $\phi_{nom,c} = 0.62$.

\begin{figure}[tb]
\centering
\includegraphics[width=0.75\textwidth]{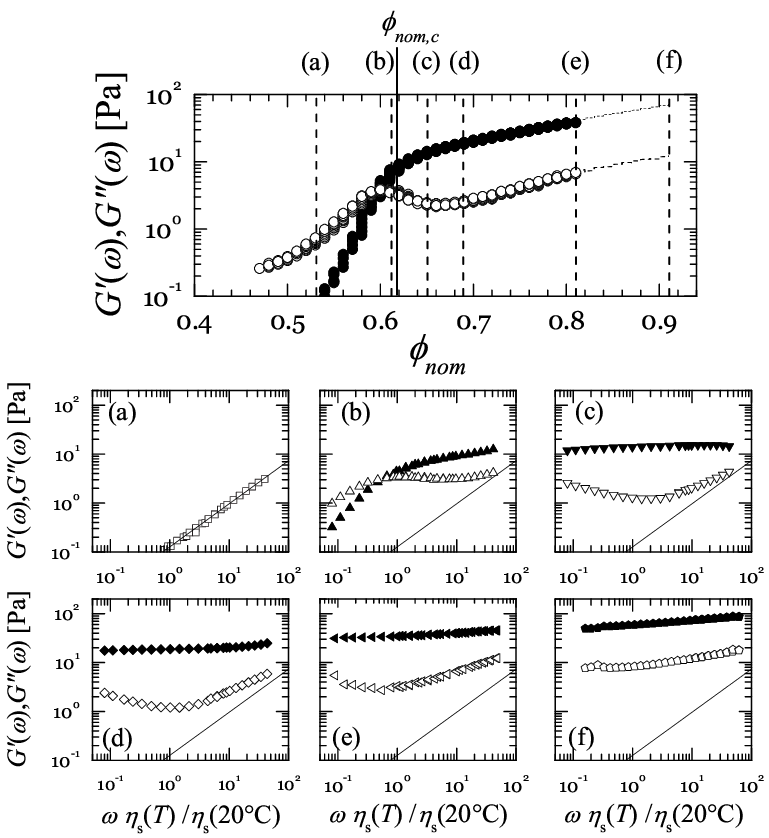}
\caption[Volume fraction dependence of $G',G''$ during a temperature
quench] {Development of storage $G'$ (solid symbols) and loss
modulus $G''$ (open symbols) during a temperature ramp from
$30\unit{^\circ C}$ to $20\unit{^\circ C}$ for a microgel system
with $c = 0.0463 \unit{g/ml}$. Additional results obtained
for a microgel system with $c = 0.0518 \unit{g/ml}$ are shown as
dashed underlying lines. The measurements are performed at a
constant frequency of $\omega = 10 \unit{rad/s}$ and a constant
strain of $\gamma =0.002$. The data are reported as a function of
the nominal volume fraction, which increases as the temperature
decreases. The solid black vertical line indicates the position of
the critical volume fraction, as determined from the critical
divergence of the low shear viscosity. The vertical dashed lines
indicate the position at which the measurements shown in (a)-(f) are
taken. Frequency dependence of $G'$ and $G''$ for: (a) $c = 0.0463
\unit{g/ml}$ at $29\unit{^\circ C}$ corresponding to $\phi_{nom} =
0.53$; (b) $c = 0.0463 \unit{g/ml}$ at $27\unit{^\circ C}$
corresponding to $\phi_{nom} =$ 0.61; (c) $c = 0.0463 \unit{g/ml}$
at $26\unit{^\circ C}$ corresponding to $\phi_{nom} =$ 0.65; (d) $c
= 0.0463 \unit{g/ml}$ at $24.5\unit{^\circ C}$ corresponding to
$\phi_{nom} =$ 0.69; (e) $c = 0.0463 \unit{g/ml}$ at $20\unit{^\circ
C}$ corresponding to $\phi_{nom} =$ 0.81; (f) $c = 0.0518
\unit{g/ml}$ at $20\unit{^\circ C}$ corresponding to $\phi_{nom} =$
0.91. The solid line in each graph corresponds to the fit of the
data shown in (a). To account for the temperature dependence of the
background viscosity the frequencies are normalised by the ratio of
the solvent viscosity to the solvent viscosity at $20\unit{^\circ
C}$.} \label{fig:pnipammoduli}
\end{figure}

For our study of the states below and above the apparent divergence of the viscosity, we choose to investigate our microgel system at a concentration of $0.0463\unit{g/ml}$; this system covers the
nominal volume fraction range $0.47 \le \phi_{nom} \le 0.81$ in the
temperature range $30\unit{^\circ C} \ge T \ge 20\unit{^\circ C}$.
To gain insight in the temperature-dependent mechanical behaviour of this
system, we perform an oscillatory shear experiment at a fixed
angular frequency ($\omega  = 10 \unit{rad/s}$) and fixed strain
($\gamma = 0.002$), ramping the temperature from $30\unit{^\circ C}$
to $20\unit{^\circ C}$ at a rate of $0.01\unit{^\circ C/s}$, where
we ensure that the strain of $\gamma = 0.002$ is well within the
linear viscoelastic regime at all temperatures investigated.
Converting concentration and temperature into nominal volume
fractions enables us to report the $\phi_{nom}$-dependence of the
loss modulus $G''$ and storage modulus $G'$ in
fig.~\ref{fig:pnipammoduli}, where we indicate the critical
condition obtained from the divergence of the low shear viscosity as
a solid vertical line.  In the volume fraction range below
$\phi_{nom,c}$, we find that $G''$ initially dominates over $G'$;
with increasing $\phi_{nom}$, $G'$ increases more quickly than $G''$
and eventually dominates over $G''$. Indeed, upon approach of the
critical conditions we expect the characteristic relaxation time to
increase dramatically. Accordingly, the characteristic frequency
denoting the cross-over from the high frequency elasticity behaviour
to the low frequency viscous behaviour systematically shifts to
lower frequencies. In a test where we probe the mechanical response
at a given frequency, we thus expect to observe a transition from a
regime in which $G''$ dominates, to a regime in which $G'$
dominates, consistent with the observed behaviour. 

That our system indeed undergoes a fluid to solid transition at $\phi_{nom,c}$ can be seen in the development of the frequency dependent responses of
this system as the temperature is decreased. At $T= 29\unit{^\circ
C}$ ($\phi_{nom} = 0.53$; fig.~\ref{fig:pnipammoduli}(a)) the
material response function is essentially determined by viscous
losses; $G''$ increases nearly linearly with frequency, the magnitude of $G''(\omega)/\omega$ is consistent with the low shear viscosity measured in steady shear experiments.
Decreasing the temperature to $T= 27\unit{^\circ C}$ ($\phi_{nom} =
0.61$; fig.~\ref{fig:pnipammoduli}(b)), the material response
function exhibits the typical features of a viscoelastic fluid; the
response is characterised by a cross-over frequency beyond which
$G'$ is the dominating modulus, while $G''$ is dominating over $G'$
at lower frequencies.  Decreasing the temperature further to $T=
26\unit{^\circ C}$ ($\phi_{nom} = 0.65$;
fig.~\ref{fig:pnipammoduli}(c)) finally leads to a response which is
typical for soft glassy materials: $G'$ is dominating over $G''$ in
the entire frequency range investigated and is nearly frequency
independent; $G''$ exhibits a minimum, indicative of some
residual slow relaxation process.

Beyond $\phi_{nom,c}$, $G'$ increases with increasing $\phi_{nom}$,
where we can take $G'$($\omega = 10\unit{rad/s})$, shown in the main
graph of fig.~\ref{fig:pnipammoduli}, as a measure of the plateau
modulus $G_{p}$.  Though the increase in $G_{p}$ does not exhibit
any further characteristic feature that may indicate a change in the
samples mechanical properties, the frequency dependence of $G'$ and
$G''$ qualitatively changes as we increase $\phi_{nom}$ beyond $\sim 0.7$.
To better capture this change, we extend our investigations to a
microgel sample with $c = 0.0518 \unit{g/ml}$. The results obtained
from the temperature ramp from $30\unit{^\circ C}$ to
$20\unit{^\circ C}$ are shown as dashed underlying lines in
fig.~\ref{fig:pnipammoduli}, demonstrating the extension of
the concentration range to $\phi_{nom} = 0.91$. By increasing the
$\phi_{nom}$ from 0.69 (fig.~\ref{fig:pnipammoduli}(d)) to 0.81
(fig.~\ref{fig:pnipammoduli}(e)) to 0.91
(fig.~\ref{fig:pnipammoduli}(f)), we observe that the difference
between $G'$ and $G''$ decreases with increasing $\phi_{nom}$; the
minimum in $G''$ gradually disappears; at $\phi_{nom} = 0.91$
$G'$ and $G''$ exhibit essentially the same frequency dependence.
Moreover, while the loss modulus in the high frequency range has a
similar magnitude in the investigated range of $0.53 \le \phi_{nom}
\le 0.69$, they significantly increase beyond $\phi_{nom}\sim 0.7$. To
show this, we report in fig.~\ref{fig:pnipammoduli}(a)-(f) the fit
describing the data at $\phi_{nom} = 0.53$, where we account for the
temperature dependence of the background viscosity by normalising
the frequencies with $\eta_{s}(T)/\eta_{s}(20\unit{^\circ C})$, the
ratio of the solvent viscosity at the experimental temperature to
the one at $20\unit{^\circ C}$.  

The change in the shape of the material response
function in conjunction with the increase of the loss modulus in the
high frequency range seemingly indicates that for $\phi_{nom} > 0.7$
the migrogels are in a different state. We tentatively interpret
this change as a hallmark of the transition to a `squeezed' state, where the particles are in direct contact. In this state, the dissipative losses at high frequencies are no longer primarily determined by the hydrodynamics of individual particles, but by the direct friction of particles against another. As the direct contact forces not only determine the friction, but also the elastic modulus, the storage and the loss moduli become more coupled to each other and decay together with frequency.

Finally, we want to point out that the observed dependence of
$G_{p}$ on $\phi_{nom}$ resembles the one of emulsions beyond random
close packing (see  fig.~\ref{fig:PuseyMason}). Thus, we seemingly
miss the range where the mechanical behaviour is defined by a strong
increase of $G_{p}$ between $\phi_{g}$ and $\phi_{j}$. Such behaviour is in agreement with results obtained for similar PNiPAM-microgels (Senff \& Richtering 1999; Senff \& Richtering 2000) and related core-shell particles (Senff \textit{et al.} 1999; Deike \textit{et al.} 2001; Le Grand \& Petekidis 2008). It can be attributed to the fact that microgels and core-shell particles --- unlike emulsion droplets --- do not have a well defined interface and in fact allow for partial interpenetration and compression (Stieger \textit{et al.} 2004; Clara Rahola 2007). Deviations from hard sphere behavior are thus expected to occur at lower volume fractions than for systems with well-defined interfaces that allow for deformation only. Indeed, the strong increase of $G_{p}$ between $\phi_{g}$ and $\phi_{j}$ has been observed for non-deformable hard sphere colloids (Le Grand \& Petekidis 2008). Thus, while emulsions still exhibit hard sphere behaviour between $\phi_{g}$ and $\phi_{j}$, the softness of microgels and core-shell particles determines the behaviour over the entire range of volume fraction above $\phi_{g}$.

\section{Spatially and temporally resolved collective dynamics}
\label{results}

We investigate the collective dynamics of our microgel system with
$c = 0.0463\unit{g/ml}$, spanning the transitional range $0.47 \le
\phi_{nom} \le 0.81$. To obtain the dynamics with temporal and
spatial resolution, we use the recently introduced photon
correlation imaging (PCIm) technique applied to a small angle light
scattering experiment as described by Duri \textit{et al.} (2009).
Coherent laser light with an \textit{in vacuo} wavelength of
$\lambda_{0} = 633\unit{nm}$ illuminates the sample at normal
incidence. We use a lens with a focal length of $f_{L}=
72.5\unit{mm}$ to image the sample onto the detector of a
charged-coupled device (CCD) camera. In contrast to normal imaging,
we place an annular aperture of radius $r_{a} = 7.2\unit{mm}$ in the
focal plane of the lens, thereby ensuring that only light scattered
within a narrow range of angles reaches the detector. The speckled
image of the sample is thus formed by the light scattered at a
single magnitude of the scattering wave vector, $q$; in our case
$q=1\unit{\mu m^{-1}}$. The magnification factor $M\simeq 1$ of the
image is determined by imaging a finely marked grid. The CCD array
comprises $633 \times 483 \unit{pixels}$, corresponding to a
rectangular observation window of approximately $5.6 \times
4.3\unit{mm}^{2}$ within the sample.

\begin{figure}[tbp]
\centering
\includegraphics[width=0.9\textwidth]{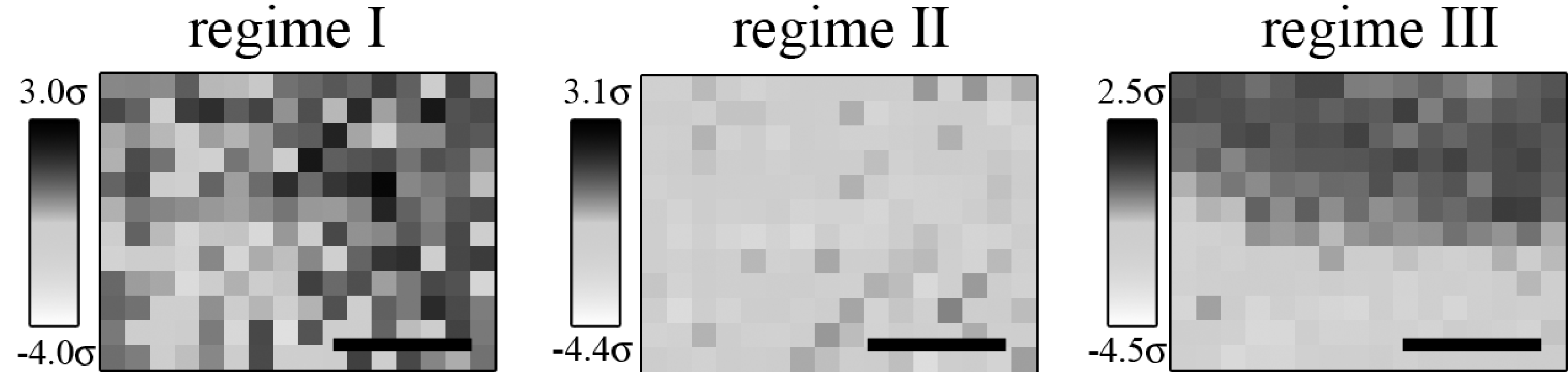}
\caption[Representative dynamic activity maps for the PNiPAM
solution] {Representative dynamic activity maps for the PNiPAM
solution quenched to regime~I ($T= 29.00\unit{^\circ C}, \phi_{nom}
= 0.53$), regime~II ($T= 24.50\unit{^\circ C}, \phi_{nom} = 0.69$),
and regime~III ($T= 20.00\unit{^\circ C}, \phi_{nom} = 0.81$). The
scale bars correspond to {2\unit{mm}}, and the intensity scale for
$c_{I}\left(t_{w},\tau,\mathbf{r}\right)$ is given in terms of its
temporal standard deviation, $\sigma$.} \label{fig:DAMs3temps}
\end{figure}

To follow the space and time-dependent fluctuations in the dynamics
of our system, we record the space-resolved speckle images in time,
where we use a camera exposure time of $2.0\unit{ms}$ and vary the
acquisition rate between $10 \unit{Hz}$ and $0.167 \unit{Hz}$
depending on the sample dynamics. Any change of the sample
configuration results in a change of the speckle pattern, which we
quantify by calculating a space and time resolved intensity
correlation function. We divide the full image into regions of
interest (ROIs) of $37 \times 37 \unit{pixels}$ corresponding to
$350 \times 350\unit{\mu m^{2}}$ within the sample. The local degree
of correlation $c_{I}\left(t_{w},\tau, \mathbf{r} \right)$ between
two images taken at time $t_{w}$ and $t_{w} +\tau$ and for a ROI
centered around the position $\mathbf{r}$ is calculated according
to:
\begin{equation}
  c_{I}\left(t_{w},\tau,\mathbf{r}\right) =
  \frac{1}{\beta}\frac{\left\langle I_{p}\left(t_{w}\right) I_{p}\left(t_{w}+\tau\right) \right
  \rangle_{ROI(\mathbf{r})}}{\left\langle I_{p}\left(t_{w}\right)\right
  \rangle_{ROI(\mathbf{r})} \left\langle I_{p}\left(t_{w}+\tau\right) \right
  \rangle_{ROI(\mathbf{r})}} -1
    \label{eq:cI}
\end{equation}
The quantity $I_{p}$ is the intensity measured at a single pixel,
$\left\langle \ldots\right\rangle_{p \in ROI(\mathbf{r})}$
represents an average over pixels belonging to a ROI centered in
$\mathbf{r}$, and the coefficient $\beta$ is a coherence factor that
depends on the speckle to pixel size ratio (Goodman 1984); it is
chosen so that $c_{I}\left(t_{w},\tau,\mathbf{r}\right) \rightarrow
1$ for $\tau \rightarrow 0$. The spatially averaged but
time-resolved degree of correlation, $c_{I}\left(t_{w},\tau\right)$,
is obtained by taking the averages over a single ROI encompassing the full image. A further average over time yields the usual intensity correlation
function $g_2(\tau)-1$ measured in traditional dynamic light
scattering: $g_2(\tau)-1 = \overline{c_{I}\left(t_{w},\tau\right)}$,
where $\overline{\ldots}$ denotes an average over $t_w$. The
spatially resolved information on the change of configuration
between $t_w$ and $t_w+\tau$ can be conveniently visualized by
constructing a ``dynamic activity map (DAM)''. Representative dynamical activity maps of our sample at
different temperatures are shown in fig.~\ref{fig:DAMs3temps}; a
physical interpretation of these maps will be given later. Each
metapixel of the dynamical activity map corresponds to a ROI and its
colour symbolises the local degree of correlation, which we indicate in units of the standard deviation of $c_{I}\left(t_{w},\tau,\mathbf{r}\right)$, $\sigma$ (Duri \textit{et al.} 2009). 

To control the temperature, we confine our microgel solution between
two sapphire windows, where a metal spacer of thickness $2\unit{mm}$
between the two windows determines the sample thickness along the
optical axis. Square thermoelectric (Peltier) heaters with a central
hole are affixed directly to the sapphire windows to serve as a heat
source. Rejected heat from the Peltier elements is removed through
contact with large aluminum blocks cooled by external water
circulation; these heat sinks have, like the Peltier elements, a
central hole placed along the optical axis to allow the light to
pass the sample.  A digital PID temperature controller controls the current output to the thermoelectric devices, where we use a $100\unit{k\Omega}$
thermistor inserted into the sample as a temperature probe.  Once
the temperature set point is reached, the control is extremely
stable; the sample temperature is recorded during the experiments
and the fluctuations rarely exceed $\pm 0.002^\circ \unit{C}$.  All
interfaces are lightly coated with a thermal grease to ensure
efficient heat transfer.

\begin{figure}[tb]
\centering
\includegraphics[width=0.8\textwidth]{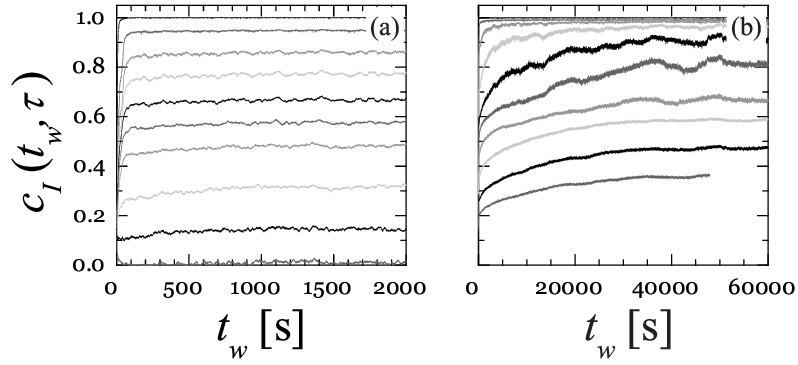}
\caption[Time evolution of $c_{I}\left(t_{w},\tau\right)$ for
$T=28.0^\circ\unit{C}$ and $T=24.5^\circ\unit{C}$] {Time evolution of the
spatially-averaged degree of correlation
$c_{I}\left(t_{w},\tau\right)$ for a temperature quench from
$T=32^\circ\unit{C}$ to (a) $T=28.00^\circ\unit{C}$ ($\phi_{nom} = 0.57$)
and to (b) $T=24.50^\circ\unit{C}$ ($\phi_{nom} = 0.69$). From top to bottom the lag times are respectively for (a) $\tau\ = 0.5$ s, $5$ s, $10$ s, $15$ s, $25$ s, $50$ s, $100$ s, $250$ s, $700$ s and $3000$ s, and for (b) $\tau\
= 3$ s, $150$ s, $600$ s, $1500$ s, $3000$ s, $5100$ s, $9000$ s,
$15000$ s, $27000$ s and $42000$ s. In both cases,
there is a transient increase in $c_{I}\left(t_{w},\tau\right)$,
which exceeds the duration of the temperature quench of 50--100~s.
The time needed to reach a quasi-stationary behaviour is
significantly longer for $\phi_{nom}>\phi_{nom,c}$ than for
$\phi_{nom}<\phi_{nom,c}$.} \label{fig:cIpnipamSALS}
\end{figure}

Prior to any experiment, our sample is equilibrated at $32^\circ
\unit{C}$ to ensure full fluidization as a starting condition.
The moment we lower $T$ to the desired experimental temperature is
defined as $t_{w} = 0$. The temperature is typically
stabilized at the set temperature within 50--100s, depending on the
temperature difference between starting and set temperature. We
follow the evolution of the dynamics of our samples by recording the
speckle pattern, subsequently calculating the spatially averaged
instantaneous degree of correlation $c_{I}\left(t_{w},\tau\right)$
at different lag times. Representative examples of the evolution
after the quench for respectively $\phi_{nom} < \phi_{nom,c}$
($28.0^\circ\unit{C}$, $\phi_{nom} = 0.57$) and $\phi_{nom} >
\phi_{nom,c}$ ($24.50^\circ\unit{C}$, $\phi_{nom} = 0.69$) are shown
in fig.~\ref{fig:cIpnipamSALS}(a) and (b). In both cases
$c_{I}\left(t_{w},\tau\right)$ increases after the quench to then
reach a quasi-stationary behaviour, indicating that the system is
characterised by a well-defined slow relaxation process for both
$\phi_{nom} < \phi_{nom,c}$  and  $\phi_{nom} > \phi_{nom,c}$.
Though this is not surprising for $\phi_{nom} < \phi_{nom,c}$, where
the system rheology is defined by a low shear viscosity (fluid-like
behaviour), it is somewhat unexpected for $\phi_{nom} >
\phi_{nom,c}$, where the system rheology is defined by a low
frequency elasticity (solid-like behaviour). Indeed, beyond
$\phi_{nom,c}$, we would in principle expect that any residual slow
dynamics would continuously evolve with $t_{w}$, exhibiting the
typical characteristics of ageing solid-like systems (van Megen
\textit{et al.} 1998; Cipelletti \textit{et al.} 2000; Bandyopadhyay
\textit{et al.} 2004).  However, as our particles are deformable we
may conceive that relaxation processes via shape fluctuations
eventually lead to the quasi-stationary dynamics observed. Though
quenches below or above $\phi_{nom,c}$ both lead to a
quasi-stationary slow dynamics, we find distinct differences in the
evolutionary behaviour between the two conditions.  While the
quasi-stationary behaviour is reached after $\approx 500\unit{s}$
for $\phi_{nom} < \phi_{nom,c}$, we typically need to wait 20000 --
30000 seconds to reach the quasi-stationary behaviour for
$\phi_{nom} > \phi_{nom,c}$.  These differences in the waiting time
seemingly reflect the differences between the situation in which the
system is quenched to a state where the individual particle
diffusion is nearly suppressed (supercooled fluid) and the one where
the system is quenched to a state where the individual particle
diffusion is suppressed (glassy system).

\begin{figure}[tb]
\centering
\includegraphics[width=0.55\textwidth]{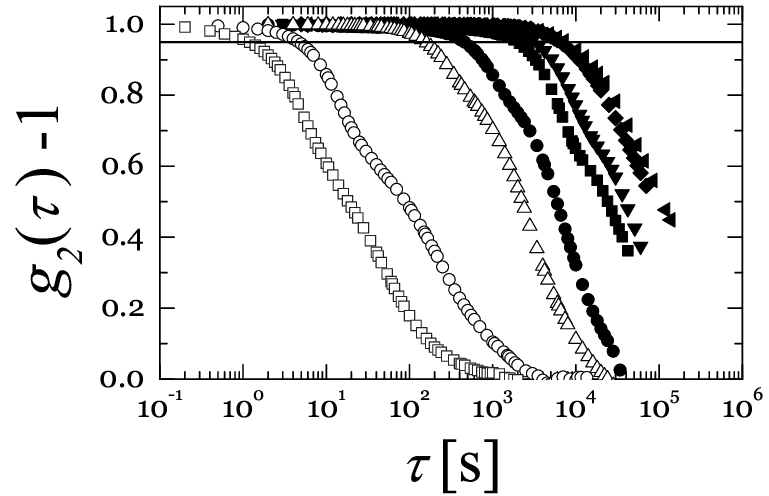}
\caption[Temperature dependence of
$\mathrm{g_{2}}\left(\tau\right)-1$ measured at low $q$] {Mean
autocorrelation function determined by taking time averages of
$c_{I}\left(t_{w},\tau\right)$ over the time windows where the
dynamics of the system is quasi-stationary. From left to right
$T=29.0, 28.0, 26.0, 25.0, 24.5, 24.0, 23.0, 20.0^\circ\unit{C}$;
accordingly $\phi_{nom} = 0.53, 0.57, 0.65, 0.68, 0.69, 0.71, 0.74,
0.81$. Open symbols denote systems that exhibit a low shear
viscosity, closed symbols denote systems that exhibit a low
frequency elastic modulus. The solid line indicates
$\mathrm{g_{2}}\left(\tau\right)-1 = 0.95$, the degree of
correlation chosen for the investigation of the temporal and spatial
fluctuations in the dynamics of our system.}
\label{fig:g2pnipamSALS}
\end{figure}

The quasi-stationary dynamics enables us to characterise the mean
dynamics of our system at different volume fractions by
determining the space and time averaged correlation functions
$\mathrm{g_{2}}\left(\tau\right)-1 =
\overline{c_{I}\left(t_{w},\tau\right)}$ over the time window where
we observe the dynamics to be quasi-stationary.  As shown in
fig.~\ref{fig:g2pnipamSALS}, $\mathrm{g_{2}}\left(\tau\right)-1$
exhibits a strong dependence on $\phi_{nom}$; the characteristic
decay times evolve from $\approx 10\unit{s}$ to $\approx
10000\unit{s}$ within the temperature range investigated. At the
lowest temperatures investigated, i.e. the largest nominal volume
fractions, the duration of our experiments is insufficient to
capture the full decay of the correlation function.  To assess the
absolute scale of $\mathrm{g_{2}}\left(\tau\right)-1$ in these
experiments we determine the baseline of
$\mathrm{g_{2}}\left(\tau\right)-1$ by calculating the degree of
correlation between speckle images taken at $32.0^\circ\unit{C}$ and
speckle images taken at the set temperature. Indeed, this
temperature jump leads to a complete reconfiguration of the system
and thus guarantees that our procedure yields the lowest possible
degree of correlation, which we use as a measure of the baseline
value. As we are working at low $q$, we presume that we capture the
entire short time relaxation spectrum within the time window accessible with
the camera, such that we normalize the intercept of our correlation
functions to one by using the $\mathrm{g_{2}}\left(\tau\right)-1$
value obtained by extrapolating the short time behaviour to $\tau =
0$.

The mean correlation functions exhibit a not well developed two-step
decay over a wide range of $\phi_{nom}$, which we attribute to
instabilities in our experimental set-up. Indeed, it is worth
recalling that our experiments are performed at low $q$, well within
the $q$-range where we probe collective diffusion. Estimating the
experimental observational length-scale as $2\pi/q = 6.3\unit{\mu
m}$, we find that we are probing a length-scale that corresponds to
$\approx 30$ particle diameters. The probed dynamics thus entails
fluctuations of the refractive index on length-scales large compared
to the particle size. We therefore do not expect to observe the
typical features of glassy dynamics probed at length scales
comparable to the particle size, where the fast local diffusion of a
particle trapped in a cage of nearest neighbours and the slow
structural rearrangement of the cage it-self lead to a two-step
decay in the intensity correlation function (van Megen \& Underwood
1993; van Megen \textit{et al.} 1998). At low $q$ the displacements of the particles diffusing within the cages of nearest neighbours are insufficient to lead to any significant dephasing of the light and thus are not contributing to the decay of the intensity correlation function.  Instead, we expect the temporal evolution of the density fluctuations to be the main cause for a
decay in $\mathrm{g_{2}}\left(\tau\right)-1$ at low $q$.

To quantify these dynamics, we choose to focus on the data obtained
at lag times where $\mathrm{g_{2}}\left(\tau\right)-1 =
\overline{c_{I}\left(t_{w},\tau\right)} = 0.95$, as marked by the
horizontal line in fig.~\ref{fig:g2pnipamSALS}. Though somewhat
arbitrary, this choice is based on the following considerations.
When analyzing the characteristic decay of a correlation function to
$\sqrt{\mathrm{g_{2}}\left(\tau\right)-1} = 1/e$, we generally
consider that the system has to reconfigure on a length scale of
$\sim 2\pi/q$.  For a significantly smaller amount of dephasing,
like the one chosen, we can think of the lengthscale over which
things have to reconfigure to be reduced.  As mentioned before,
$2\pi/q$ is in our experiment rather large, corresponding to
$\approx 30$ particle diameters; by monitoring the fluctuations in
$c_{I}$ with a mean of $\overline{c_{I}\left(t_{w},\tau\right)} =
0.95$, we expect to resolve heterogeneities in collective
rearrangements on lengthscales smaller than 30 particle diameters.
Moreover, in order to have sufficient statistics in
$c_{I}\left(t_{w},\tau\right)$, the system has to reconfigure
several times over the chosen observational lengthscale. For the
largest $\phi_{nom}$ investigated, the time to reconfigure so that
$\overline{c_{I}\left(t_{w},\tau\right)} = 0.95$ is already of the
order of $10^{5}\unit{s}$, near to the duration of the experiment;
here, the processing of a lower degree of correlation and thus
larger lag time is precluded because of the finite duration of our
experiment.  Finally, the choice of a lag time which is
significantly smaller than the characteristic decay time enables us
to use the direct noise correction scheme described by Duri
\textit{et al.} (2005, Sec. IVc).

\begin{figure}[tbp]
\centering
\includegraphics[width=0.7\textwidth]{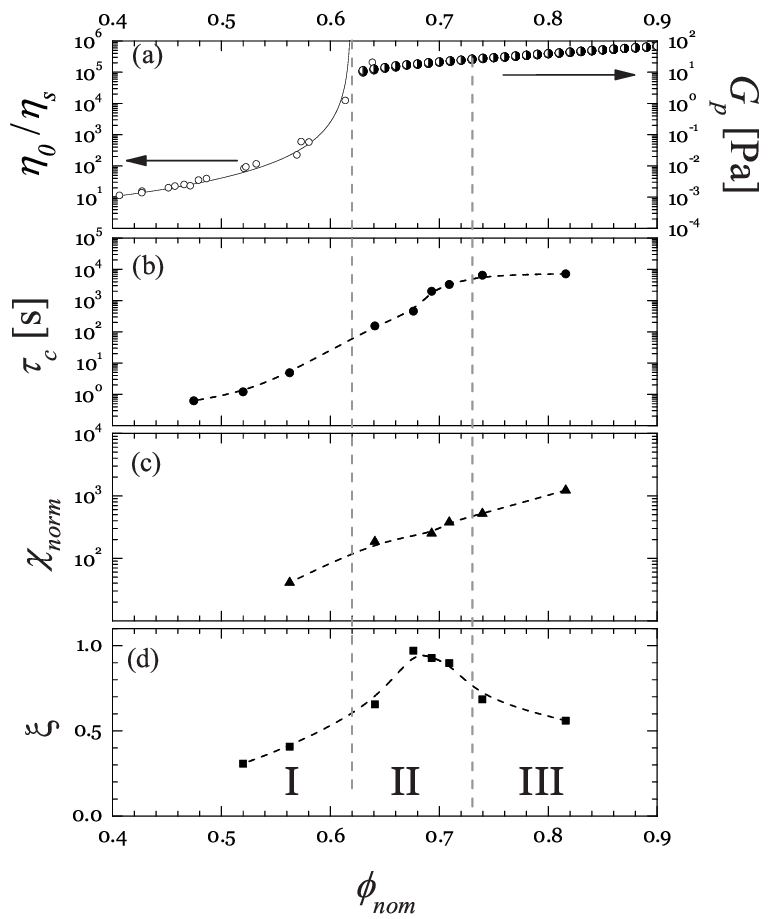}
\caption[Volume fraction $\phi_{nom}$ dependence of the rheological
properties, relaxation time, and the temporal and spatial dynamic
heterogeneities] {Volume fraction dependence of mechanical and
dynamic properties, divided into 3 distinct regimes (\textsc{i},
\textsc{ii}, and~\textsc{iii}).  (a) Low shear viscosity (open
circles) and plateau modulus $G_{p}$ (half-filled circles). The
apparent divergence of the low shear viscosity clearly defines the
boundary between regime~\textsc{i} and~\textsc{ii}.  (b) The decay
time $\tau_{c}$ increases steadily up to $\phi_{nom}\approx 0.73$,
after which it becomes approximately constant (regime~\textsc{iii}).
(c) The normalised temporal variance $\chi_{norm}$ monotonically
increases with no distinct features over the entire investigated
range of $\phi_{nom}$. (d) The range of spatial correlations of the
dynamics, $\xi$ (see text for definition) increases with
$\phi_{nom}$ in regime~\textsc{i}, peaks in regime~\textsc{ii} and
finally decreases for highly squeezed states, regime~\textsc{iii}.
The lines serve as guides for the eye.}
\label{fig:4partPNiPAMphievolution}
\end{figure}

From our $c_{I}$-data with a mean of
$\overline{c_{I}\left(t_{w},\tau\right)} = 0.95$, we extract three
quantities that respectively characterise the average dynamics, its
temporal fluctuations, and the spatial correlations in the dynamics
of our system. The dependence of these quantities on $\phi_{nom}$ is
shown in fig.~\ref{fig:4partPNiPAMphievolution} along with the
mechanical characteristics, which we use as a gauge of the
transitional behaviour. As a parameter characterising the average
dynamics, we determine the lag time at which
$\overline{c_{I}\left(t_{w},\tau\right)} = 0.95$, $\tau_{c}$. Though
$\tau_{c}$ significantly increases as $\phi_{nom}$ is increased, the
fluid-solid transition, as defined by the mechanical properties of
our system, is not reflected by any pronounced feature in the
$\phi_{nom}$-dependence of $\tau_{c}$.  For the volume fraction
range below $\phi_{nom,c}$, the characteristic time strongly
increases, while at the transition this increase is somewhat slowed
down; for $\phi_{nom} > 0.73$, $\tau_{c}$ reaches a
constant value.

To characterise the temporal fluctuations in the dynamics of our
system, we calculate $\chi_t$, the temporal variance of
$c_{I}\left(t_{w},\tau\right)$, again fixing $\tau = \tau_c$ such
that $\overline{c_{I}\left(t_{w},\tau_c\right)} = 0.95$.  We recall
that $c_{I}\left(t_{w},\tau\right)$ is the instantaneous degree of
correlation obtained by taking the pixel average in
Eq.~(\ref{eq:cI}) over the entire image. Thus, $\chi_t$ quantifies
the temporal fluctuations of the spatially averaged dynamics,
similarly to the dynamical susceptibility $\chi_4$ introduced in
numerical works on glassy systems (La\v{c}evi\'{c} 2003). To gauge
the significance of $\chi_t$ with respect to the noise stemming from
the finite number of speckles recorded, we normalise our data with
data obtained in reference measurements, where we use freely
diffusing colloidal particles as a model system exhibiting
homogeneous dynamics. For the Brownian particles, $\chi_t$ contains
only the noise contribution. Any excess of the variance with respect
to the value obtained for the Brownian particles can thus be
ascribed to temporally heterogeneous dynamics. For all $\phi_{nom}$
investigated, the normalised temporal variance $\chi_{norm}$ is
significantly larger than the one expected for homogeneous dynamics,
as shown in fig.~\ref{fig:4partPNiPAMphievolution}(c); this
indicates that the dynamics of our system is characterised by
heterogeneities at all volume fractions investigated, temporal
heterogeneity becoming more pronounced as the volume fraction
increases, as denoted by the increase of $\chi_{norm}$ with
$\phi_{nom}$.

As discussed for instance in Trappe \textit{et al.} (2007),
$\chi_{norm}$ depends on both the amount of temporal fluctuations of
the dynamics at a given location and the range of spatial
correlations of the dynamics. To characterise the spatial
fluctuations in the dynamics of our system, we process the speckle
images calculating the degree of correlation with spatial
resolution, $c_I(t_w,\tau,\textbf{r})$. By contrast to the mean dynamics and
the temporal variation of the dynamics, the dynamic activity maps
obtained from a space-resolved analysis at a lag $\tau = \tau_c$
reveal striking differences depending on whether $\phi_{nom}$ is
below $\phi_{nom,c}$, just above $\phi_{nom,c}$ or above $\phi_{nom}
\approx 0.73$,  where we identify the three conditions as regime~I,
regime~II, and regime~III. At $\phi_{nom} = 0.53$ (regime~I), the
dynamics are characterised by local fluctuations of
$c_I(t_w,\tau,\textbf{r})$, which can be visually assessed from the
fluctuations of the intensity in the dynamical activity map shown in fig.~\ref{fig:DAMs3temps} (left image). These fluctuations appear to be weakly correlated.  At $\phi_{nom} = 0.69$ (regime~II), the DAMs
have, at any given time $t_w$, essentially the same intensity level
in all metapixels, as shown in fig.~\ref{fig:DAMs3temps} (middle image). This 
suggests that the dynamics are correlated over
distances comparable to the system size. At $\phi_{nom} = 0.81$
(regime~III), we find large variations in the DAM intensity, with
regions of high dynamical activity coexisting with ``quieter''
zones, as shown in fig.~\ref{fig:DAMs3temps} (right image). The boundary between the high and low dynamical activity
zones appears to be well defined; these zones typically extend over
a sizeable fraction of the field of view, but do not extend over the entire field of view like in regime~II. As a further difference between the three
regimes, we note that the temporal evolution of the spatial
heterogeneities strongly depends on whether we quench the system to
regime~I,~II, or~III. While for regime~I and~II the dynamical
activity within one region quickly switches from low to high and
vice versa, the dynamic activity within a given zone persists for a
long time in regime~III.

\begin{figure}[tbp]
\centering
\includegraphics[width=0.55\textwidth]{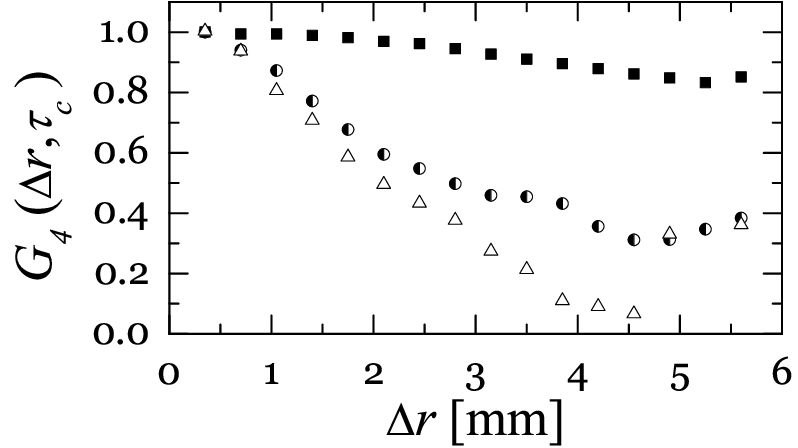}
\caption[Spatial correlation, $G_4(\Delta r,\tau_c)$, of the
dynamics] {Spatial correlation of the dynamics, $G_4(\Delta
r,\tau_c)$, as defined in the text. The range of spatial
correlations of the dynamics is the smallest in regime~I (triangles), becomes comparable to the system size in
regime~II (squares), and finally decreases in regime~III
(semi-filled circles).} \label{fig:G4}
\end{figure}

To quantify the spatial correlations of the dynamics, we calculate
the four-point correlation function $G_4(\Delta r,\tau)$ introduced
by Duri \textit{et al.} (2009). This function compares the local
dynamics on a time scale $\tau$ in regions separated by a distance
$\Delta r$. Its definition is similar to that used in numerical work
(La\v{c}evi\'{c} \textit{et al.} 2003):
\begin{equation}
G_4(\Delta r,\tau) = B \left \langle \overline{\delta
c_I(t_w,\tau,\mathbf{r}) \delta c_I(t_w,\tau,\mathbf{r'})} \right
\rangle_{\Delta r} \, , \label{eq:G4}
\end{equation}
where $\delta c_I(t_w,\tau,\mathbf{r}) = c_I(t_w,\tau,\mathbf{r}) -
 \overline{c_I(t_w,\tau,\mathbf{r})}$, $B$ is a constant such that
$G_4(\Delta r, \tau) \rightarrow 1$ for $\Delta r \rightarrow 0$,
and $\left\langle\ldots\right\rangle_{\Delta r}$ is the average over
all pairs of $\mathbf{r}$ and $\mathbf{r'}$ corresponding to the
same distance, $\Delta r = \left|\mathbf{r}-\mathbf{r'}\right|$. As
for the other dynamical quantities reported in
fig.~\ref{fig:4partPNiPAMphievolution}, we analyse $G_4$ for $\tau =
\tau_c$. The spatial correlation functions corresponding to the
three regimes shown in  fig.~\ref{fig:DAMs3temps} are shown in
fig.~\ref{fig:G4}.

In regime~I, $G_4$ decays over a few metapixels, corresponding to
about 2 mm. By contrast,  in regime~II $G_4 \approx 1$ over the full
field of view, implying that the dynamics are almost perfectly
correlated over distances at least as large as $5\unit{mm}$. In regime~III, the range of spatial correlations is
reduced, although it remains higher than in regime~I. In order to
extract a characteristic length describing the range of correlated
dynamics, in principle one would fit $G_4$ to some
functional form, such as an exponential. While this approach would
be sound at the lowest and highest volume fractions investigated, it
would fail in regime~II, because $G_4$ is essentially flat over the
full accessible range of spatial delays. Instead, we introduce a
normalized correlation length by defining
\begin{equation}
\xi = \Delta r_{max}^{-1} \int _{0}^{\Delta r_{max}} G_4(\Delta r,
\tau_c) \mathrm{d}\Delta r \label{eq:xi} \, ,
\end{equation}
where $\Delta r_{max} = 5.6$ mm is the largest spatial delay
accessible in our experiment. This correlation length varies between
0 if the dynamics of distinct metapixel are totally uncorrelated and
1 if the dynamics are perfectly correlated over the full field of
view. Reflecting the trend shown by $G_4$ in fig.~\ref{fig:G4},
$\xi$ starts at a low level at low $\phi_{nom}$, increases
dramatically up to almost unity in regime~II and finally decreases
in regime~III, as shown in
fig.~\ref{fig:4partPNiPAMphievolution}(d).

Consistent with the growth of $\xi$ observed here in
regime~I and~II, confocal microscopy (Weeks \textit{et al.} 2007) and dynamic light scattering experiments (Berthier \textit{et al.} 2005; Brambilla
\textit{et al.} 2009) have shown that for colloidal hard spheres the
range of correlated dynamics continuously increases with volume
fraction, this increase extending beyond the apparent divergence of
the structural relaxation time at the experimental glass transition. 
However, the correlation length measured here beyond $\phi_{g}$, which we identify as the volume fraction at which the low shear viscosity apparently diverges, is much larger than the one reported for the self-diffusion of hard spheres, where the dynamics are correlated only up to a few particle sizes (Weeks \textit{et al.} 2007). In this context, it is worth stressing that we do not probe the dynamics on a particle level. Indeed, our experiment is performed at low $q$ and a given ROI extends over nearly 2000 particles. Our experiment thus probes the reconfiguration of large spatial density fluctuations and the correlation length measured indicates how zones of higher or lower dynamical activity are correlated with each other; this does not necessarily correspond to the actual number of cooperatively rearranging particles that can be measured in real space experiments. Indeed, spatial correlations that span the whole sample have recently been inferred in a study of the dynamical heterogeneities of xenospheres by diffusing wave spectroscopy (Ballesta \textit{et al.} (2008)). In this study, Ballesta \textit{et al.} showed that the correlation length grows with increasing volume fracion to eventually span the whole sample near the maximal packing conditions, in agreement with the behavior observed in regime~I and~II. For the xenospheres, measurements beyond the maximal packing condition were not possible as the xenospheres are essentially non-deformable. The striking decrease of $\xi$ at volume fractions beyond $\phi_{j}$ observed in our system is instead reminiscent of the behavior reported for a 2-dimensional driven granular system (Lechenault \textit{et al.} 2008), where a non-monotonic $\phi$ dependence in the spatial corrrelation of the dynamical heterogeneities has been observed as well. 

Together with these suggestive analogies, our data can be rationalized considering the transition from a situation where the particles are densely packed but do not exert direct contact forces on one another (regime~I and II) to a situation where they do (regime~III). In regime I (supercooled state) and II (glassy state) the dynamics is purely determined by thermal motion. With increasing volume fraction, the dynamics becomes increasingly cooperative, reaching in regime II  a situation where any dynamical activity at a given zone will require the neighbouring zone to be dynamically active as well. By contrast, a quench into regime~III (squeezed state) imposes large deformations on the microgels. The dynamic activity observed in this regime is therefore likely to be at least partly due to imbalanced stresses; these lead to rearrangements, which do not appear to equilibrate the stress imbalances.  As a result of this the dynamical activity persists in time and elastically propagates in space, albeit remaining well localized,  exhibiting sharp boundaries to dynamically inactive zones. 

Finally, we note that the almost complete lack of hallmarks in $\chi_{norm}$ at
$\phi_{j}$ is somewhat surprising. Indeed, we generally would expect that $\chi_{norm}$ becomes maximal when $\xi$ is maximal.  The increase of $\chi_{norm}$ beyond $\phi_{j}$ indicates that the amplitude in the dynamical fluctuations of the spatially averaged signal $c_{I}\left(t_{w},\tau\right)$ dramatically increases beyond $\phi_{j}$,  thereby compensating the effect of the decreasing correlation length. A full understanding of this effect in conjunction with the $\phi$ dependence of $\tau_{c}$ is the subject of further research.

\section{Conclusions}
We have characterized the mechanical and dynamical properties of a
PNiPAM-microgel system whose dimensions are conveniently varied by
temperature.  Despite the fact that these microgels do not possess a
well-defined interface, allow for partial interpenetration and
compression, a remarkable number of the mechanical features of hard
and soft, deformable spheres with well-defined interfaces are
reproduced. The low shear viscosity appears to critically diverge at
some nominal critical volume fraction $\phi_{nom,c}$, which we
define as the transition from a `supercooled' to a `glassy' state. As the volume fraction of our microgel system is ill-defined, the
indication in terms of nominal volume fractions does not allow for
any conclusions with respect to the exact volume fraction at this
transition. Moreover, previous work has shown that the lower density
shells of our microgels partly interpenetrate near $\phi_{nom,c}$
(Clara-Rahola 2007), which adds to the difficulty in precisely
assessing the fluid to solid transition observed at $\phi_{nom,c}$.
The indication `supercooled' and `glassy' here refer to the inferred
conditions that the microgels are able to escape out of a cage of
nearest neighbours in the `supercooled' state, while this is not the
case in the `glassy' state, where we additionally presume that the
microgels do not exert direct contact forces on each other in the
`glassy' state. Despite the evident disadvantages in using microgels
to investigate the fluid-solid transitions of repulsive systems, the
advantage of varying the volume fraction by changing the temperature
enables us to precisely control the history of our sample, which we
believe prevails over the disadvantages. All our experiments are
performed by first equilibrating the system in the fluid state at
high temperature, which is then quenched to a given solid state by
lowering the temperature. The residual dynamics in the solid-like
states are thus not affected by the shear history of the sample
other than the one imposed by the inflation of the particles.

We measure this residual dynamics at a $q$-vector where we probe the
collective diffusion of our system. Approaching the critical nominal
volume fraction from below, we find that the extent of the spatial
correlation in the collective dynamics increases, reflecting the
increasing constraints set by the increasing volume fraction, whereby 
a local reconfiguration of the system can only occur
when the neighbouring areas also reconfigure. Beyond
$\phi_{nom,c}$ the range of spatial correlations extends over the
entire observational window. The mechanical properties are
characterised by a low frequency elasticity; the dissipative losses
probed at high frequencies indicates that the local dynamics of the
particles are still determined by the hydrodynamics of individual
particles.  Increasing the volume fraction further, we identify a
second transition, the transition to the `squeezed' states.  In
these states we find that the storage and dissipative contributions
in the mechanical properties of the system become increasingly
coupled and that the high frequency dissipative losses become
significantly larger than the ones observed at lower $\phi_{nom}$. We
attribute this behaviour to friction between the particles
resulting from the fact that the particles exert direct forces on
one another.  In this state, the dynamics exhibit spatial
heterogeneities that are characterised by large zones of
respectively high and low dynamical activity.  This behaviour
appears to be a hallmark of squeezed systems, where the quench into
this state does not necessarily lead to a balanced-stress situation.
This imbalance leads to rearrangements, which again may or may not
result in a balanced-stress situation, such that the dynamical
activity persists in time.  The ranges of the dynamically correlated
regions here are smaller than in the `glassy' state, indicating that
stress-driven rearrangements are more localized than the
thermally-driven rearrangements in the glassy state. To our knowledge, our data are the first demonstrating the intrinsic changes in the dynamical heterogeneities of deformable colloidal systems as the
origin of the dynamical process switches from purely thermal to
stress-driven.

\begin{acknowledgements}
We gratefully acknowledge financial support from the Swiss National
Science Foundation (grant no.\ 200020-117755 and 200020-120313), CNES
and  CNRS (PICS no.\ 2110). L.\ C.\ acknowledges support from the
Institut Universitaire de France.

\end{acknowledgements}

\label{lastpage}
\end{document}